# Study of electromagnetic enhancement for surface enhanced Raman spectroscopy of SiC graphene

Jing Niu[1], Viet Giang Truong[1], Han Huang[2,3], Sudhiranjan Tripathy[4], Caiyu Qiu[5], Andrew T.S. Wee[2,3], Ting Yu[2,3,5], and Hyunsoo Yang[1,3,a]

[1]Department of Electrical and Computer Engineering, National University of Singapore, 117576, Singapore

[2]Department of Physics, National University of Singapore, 117542, Singapore

[3]Graphene Research Centre, National University of Singapore, 117546, Singapore

[4] Institute of Materials Research and Engineering, 117602, Singapore

[5] Division of Physics and Applied Physics, School of Physical and Mathematical Sciences, Nanyang Technological University, 637371 Singapore, Singapore

The electromagnetic enhancement for surface enhanced Raman spectroscopy (SERS) of graphene is studied by inserting a layer of $Al_2O_3$ between epitaxial graphene and Au nanoparticles. Different excitation lasers are utilized to study the relationship between laser wavelength and SERS. The theoretical calculation shows that the extinction spectrum of Au nanoparticles is modulated by the presence of graphene. The experimental results of the relationship between the excitation laser wavelength and the enhancement factor fit well with the calculated results. An exponential relationship is observed between the enhancement factor and the thickness of the spacer layer.

[a] E-mail address: eleyang@nus.edu.sg

Since graphene was made by mechanical exfoliation in 2004, it has stirred enormous interest from various research fields due to its extraordinary properties including a high electron mobility of up to 200,000 cm$^2$/Vs, large thermal conductivity exceeding 3,000 W/mK at room temperature, and low opacity of ~2.3%.[1-8] The properties of graphene can be altered largely by its fine structure, for example, through the number of layers, doping level, and defects.[9-11] Raman spectroscopy is a common and non-destructive tool to characterize the structural characteristics of graphene. The number of layers of graphene can be easily distinguished by the evolution of the shape of the 2$D$ peak and the shifting of the peak position of both the 2$D$ and $G$ peaks.[12] It is also capable of monitoring the doping, defects, thermal conductivity, and strain of graphene.[13-16] However, the Raman signal of graphene is not strong enough for some fine structural characteristics such as vacancies, edge structures, and crumpling.[17] This is mainly due to the fact that for Raman scattering, the majority of the incident light undergoes Rayleigh scattering which does not contribute to the Raman signal, and the greater part of the incident light is transmitted through graphene without contributing to the scattered radiation.[17] Therefore, researchers are continuously working on new methods to improve the Raman signal strength.

The substrate commonly used for mechanical exfoliated graphene is Si substrate capped with a layer of SiO$_2$ with a specific thickness, which is designed for easier observation of exfoliated graphene under optical microscopy. It is found that this special substrate structure not only enhances the optical contrast between graphene and the background, but also provides an enhancement of the Raman signal through interference-enhanced Raman scattering (IERS).[18] Hence, exfoliated graphene with the special Si/SiO$_2$ substrate always has a stronger Raman signal compared with epitaxial graphene on SiC substrate. Epitaxial graphene on SiC substrate is a promising candidate for electronic applications, since it provides a semiconducting substrate by default and comprises a uniform layer compared to chemical vapor deposition (CVD) graphene.[2] In addition to IERS, surface enhanced Raman spectroscopy (SERS) is another method that boosts the strength of the Raman signal.[19, 20] Thus far, most of the research studies about SERS of graphene have been carried out on exfoliated graphene, whereas no results have been published for epitaxial graphene on SiC substrate. For SERS, metal nanoparticles are often included in the system to induce the primary mechanisms of the enhancement which is the localized electromagnetic field surrounding the nanoparticles caused by localized surface plasmon resonances (LSPR).[21] Two main mechanisms of SERS are chemical enhancement, which requires a direct contact between graphene and nanoparticles, and electromagnetic enhancement.[22-24]

In this letter, a layer of Al$_2$O$_3$ with various thicknesses is inserted between graphene and Au nanoparticles to study the electromagnetic enhancement of SERS solely by eliminating the possibility of chemical enhancement of SERS caused by charge transfer. In addition, through variation of the thickness of Al$_2$O$_3$ the thickness effect of the spacer layer on SERS can be studied. The experimental result of the enhancement factor as a function of the excitation laser wavelength agrees well with the theoretical calculation of the extinction spectrum of LSPR, providing a selection guide to laser wavelengths in order to obtain the maximum SERS of graphene. Furthermore, an exponential relationship between the thickness of the Al$_2$O$_3$ insulating layer and the electromagnetic enhancement factor is observed.

The graphene used in our experiment is single layer graphene thin films grown on a Si-terminated 6H-SiC (0001) surface. The film thickness is confirmed by scanning tunneling microscopy (STM) followed by Raman spectroscopy.[22, 25, 26] A layer of Al with a thickness of ~3 nm was deposited on top of the sample by an electron beam evaporator followed by naturally-occurring oxidation in an ambient environment. The thickness of the thin layer is monitored by a quartz crystal during the evaporation process and is confirmed by an ellipsometer. For a thicker $Al_2O_3$ film, the above steps are repeated. Metal nanoparticles are formed by evaporating a thin Au film of 5 nm, followed by annealing in 500 °C under an $N_2$ environment for 20 min. The fabrication processes are shown in Fig. 1(a). The scanning electron microscope (SEM) images of metal nanoparticles formed on top of the sample with different $Al_2O_3$ thicknesses show no noticeable change in the average size of diameter (~15 nm). The inset of Fig. 1(b) shows one of the SEM images of nanoparticles formed on top of the sample after annealing.

The transmission spectrum shown in Fig. 1(b) is obtained to verify the excitation of localized surface plasmons in the structure. The transmission spectra without and with Au nanoparticles show apparent differences. The transmission data with Au nanoparticles shows an additional valley compared with that without nanoparticles. The minimum transmission appears at ~560 nm which matches well with the LSPR wavelength of gold nanoparticles.[27]

Raman spectra of the same sample are taken with four excitation lasers with different wavelengths (325, 488, 532, and 785 nm) on the area without and with nanoparticles, as shown in Figs. 2(a) and 2(b), respectively. The thickness of $Al_2O_3$ is 3 nm. The enhancement factor is calculated through two steps. First, the intensity of the G peak (~1580 cm$^{-1}$) for each spectrum is normalized with the SiC substrate peak at ~ 1519 cm$^{-1}$. The SiC substrate peak is comparable for samples deposited without and with nanoparticles under the same experimental conditions on the same sample, thus this step allows us to get a more reliable comparison result. Then for each wavelength, the enhancement factor is computed by using the intensity of the G peak obtained with Au particles divided by that without particles. It is observed that the enhancement factor of G peak is always greater than 1, when the sample is capped with Au particles. Figure 2(c) shows the enhancement factors of four different excitation lasers, indicating the maximum enhancement with a 532 nm laser. Previously, detailed wavelength-scanned surface-enhanced Raman excitation spectroscopy (WS-SERES) and plasmon-sampled surface-enhanced Raman excitation spectroscopy demonstrated that the maximum SERS enhancement occurs when the excitation wavelength of the laser is slightly blue-shifted from the resonance wavelength of LSPR.[28-30] Considering the above observation and that 532 nm is the nearest laser wavelength to the surface plasmon resonance wavelength (560 nm) of Au particles, strong electromagnetic fields surrounding the Au particles induce more electromagnetic enhancement to the Raman intensity. However, an unexpectedly low enhancement factor is obtained, when the 488 nm laser is utilized. As 488 nm is rather close to the LSPR wavelength, its enhancement should be higher than the one obtained with either the 325 or 785 nm laser.

It is known through WS-SERES that the line shape of SERS should be similar to that of the LSPR extinction spectra.[28, 31] In order to understand the above observation, theoretical calculations of the extinction efficiency based on dipole approximation have been carried out. In the calculation, the substrate is assumed to be flat with semi-infinite thickness and an Au nanoparticle with a diameter of 15 nm is embedded in a mixture of Al$_2$O$_3$ (30%) and air (70%) above graphene. The dielectric constant of graphene is calculated as $\varepsilon_3 = 1 + i\sigma_g / \omega\varepsilon_0 d_g$, where $\varepsilon_0$, $\varepsilon_3$, $\sigma_g$, and $d_g$ is the permittivity of vacuum, the dielectric constant, optical sheet conductivity, and thickness of graphene (0.34 nm), respectively. The calculation of the dielectric constant of graphene is based on an assumption that the optical response of the graphene layer is given by optical sheet conductivity.[32] The bulk dielectric constant of gold and Al$_2$O$_3$ is obtained from the literature.[33] The polarizability of a metal nanoparticle above a conductive substrate can be written as:

$$\alpha = 4\pi a^3 \left(\frac{\varepsilon_1 - \varepsilon_2}{\varepsilon_1 + 2\varepsilon_2}\right)\left[1 - \beta\left(\frac{a}{2d}\right)^3\left(\frac{\varepsilon_1 - \varepsilon_2}{\varepsilon_1 + 2\varepsilon_2}\right)\left(\frac{\varepsilon_3 - \varepsilon_2}{\varepsilon_3 + \varepsilon_2}\right)\right]^{-1} \quad (1)$$

where $\alpha$ is the polarizability, $a$ is the radius of the sphere (7.5 nm), $\beta$ is taken as 1 for lateral electric fields, and $d$ is the distance from the center of the gold sphere to the substrate surface.[34] $\varepsilon_1$, $\varepsilon_2$, and $\varepsilon_3$ are the dielectric constants of gold, surrounding media, and the substrate, respectively. The absorption efficiency $Q_{abs}$ is given by $Q_{abs} = [k/\pi a^2]\text{Im}(\alpha)$ and the scattering efficiency $Q_{sca}$ is given by $Q_{sca} = [k^4 / 6\pi^2 a^2]|\alpha|^2$.[34] The summation of these two gives the extinction efficiency $Q_{ext}$ which is plotted in Fig. 2(c). It is clear that the calculation result fits qualitatively well with the experimental data. This shows that the presence of graphene can change the extinction spectrum of Au nanoparticles substantially, which leads to a nonmonotonic change in the enhancement factor of SERS with different excitation wavelength lasers.

The enhancement factor can be improved by optimizing the size and interspaces of the nanoparticles.

The relationship between the enhancement factor and the distance between Au nanoparticles and graphene has been also investigated. For different $Al_2O_3$ thicknesses, the Raman data before and after the formation of Au nanoparticles have been measured. The excitation wavelength of the laser is 488 nm. Without Au nanoparticles in the structure, the intensity of the $G$ peak does not change with various $Al_2O_3$ thicknesses as shown in Fig. 3(a). After forming Au nanoparticles, the enhancement factor of the $G$ peak decreases with an increase in the $Al_2O_3$ thickness, as indicated in Fig. 3(b). The enhancement factor falls off exponentially with an increase in the insulator thickness, as shown in Fig 3(c). An enhancement factor of 1.2 is still present when the thickness of $Al_2O_3$ increases to 12 nm. The extinction efficiency is calculated with various $Al_2O_3$ thicknesses based on Eq. (1) as shown in Fig. 3(c), and it shows a decreasing trend similar to the experimental data. The observed data can be better accounted for by the exponential decay of electromagnetic fields of localized surface plasmon away from its excitation source (Au nanoparticles) because of the evanescent nature of the LSPR.[35] One recent report also showed an exponential decrease of the photoluminescence enhancement ratio as the thickness of the $SiO_2$ spacer between graphene and the ZnO layer increases.[36]

In order to ensure that there is no diffusion of Au nanoparticles through the $Al_2O_3$ spacer layer and get in contact with the graphene layer which will allow charge transfer between Au nanoparticles and graphene, Au nanoparticles are etched away using aqua regia and Raman spectra of the samples are taken again as shown in the inset of Fig. 3(c).

As shown in the spectra, the enhancement of *G* peak is absent after Au nanoparticles are etched away. This confirms that the presence of the spacer layer eliminates the possibility of any chemical enhancement of SERS caused by charge transfer. In addition, the change in the thickness of an insulator layer also affects the scattering of light and the modulation of the surface plasmon resonance wavelength[37] due to graphene could change the enhancement factor of the Raman intensity. This result can be applied to the research field where graphene is utilized as a substrate to obtain more strong Raman signals in the characterization of low concentration molecules. Experimental results show that graphene can be considered as a substrate for various molecules to obtain better Raman spectroscopy.[38, 39] Recently Au thin films on graphene have been also used to facilitate enhanced Raman signals of molecules.[40] Our proposed structure with an insulating layer will be particularly useful for certain experiments in which a direct contact between a conducting layer and molecules are not preferred.

In conclusion, we have studied the SERS induced by metal nanoparticles on SiC epitaxial graphene. By inserting a layer of $Al_2O_3$ insulator between graphene and nanoparticles, the chemical enhancement of Raman spectroscopy caused by charge transfer is ruled out. The enhancement of Raman scattering is due to the near-field plasmonic effects induced by Au nanoparticles. Excitation lasers with different wavelengths have been utilized to study the relationship between the laser wavelength and SERS. Theoretical calculations show that the extinction spectrum of Au nanoparticles is modulated by graphene. The experimental results of the relationship between the excitation laser wavelength and the enhancement factor match well with the calculated results. An exponential relationship is observed between the enhancement factor and the thickness of the Al$_2$O$_3$ spacer layer.

This work is partially supported by the Singapore National Research Foundation under CRP Award No. NRF-CRP 4-2008-06 and "Novel 2D materials with tailored properties: beyond graphene" (R-144-000-295-281).

Figure captions

Figure 1. (a) Schematic illustration of sample fabrication process. (b) The transmission spectra of the area without and with Au nanoparticles on the same sample (inset: SEM image of Au nanoparticles formed on top of $Al_2O_3$).

Figure 2. Raman spectra obtained on the same sample with different exitation laser wavelengths (325, 488, 532, and 785 nm) without (a), and with (b) nanoparticles. The thickness of $Al_2O_3$ is 3 nm. (c) Enhancement factor of the G peak and the calculated extinction efficiency of Au nanoparticles vs. wavelength of excitation lasers (inset: schemtic illustration of the structure used for therotical calculation).

Figure 3. Raman spectra of SiC graphene capped with different thicknesses of $Al_2O_3$ from samples without (a), and with (b) nanoparticles. (c) The relationship between the enhancement factor and the thickness of the $Al_2O_3$ spacer layer. The calculated extinction efficiency is shown as well. The excitation wavelength of the laser is 488 nm. The y-axis in (c) is on a logarithmic scale. (inset: Raman spectra of SiC graphene samples with different thicknesses of $Al_2O_3$ after removing Au nanoptraticles with aqua regia).

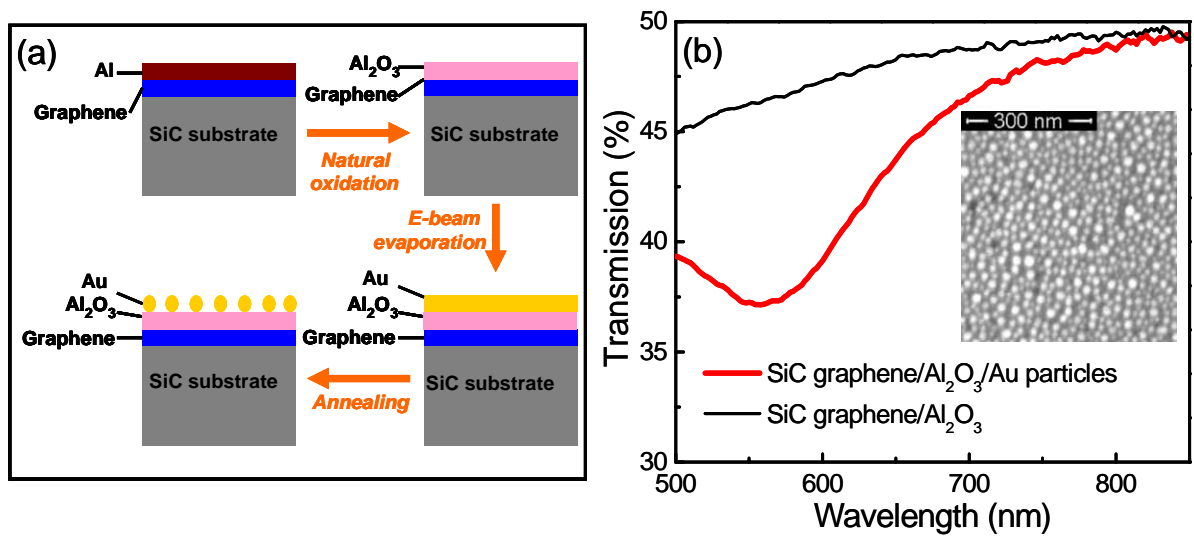

Figure 1

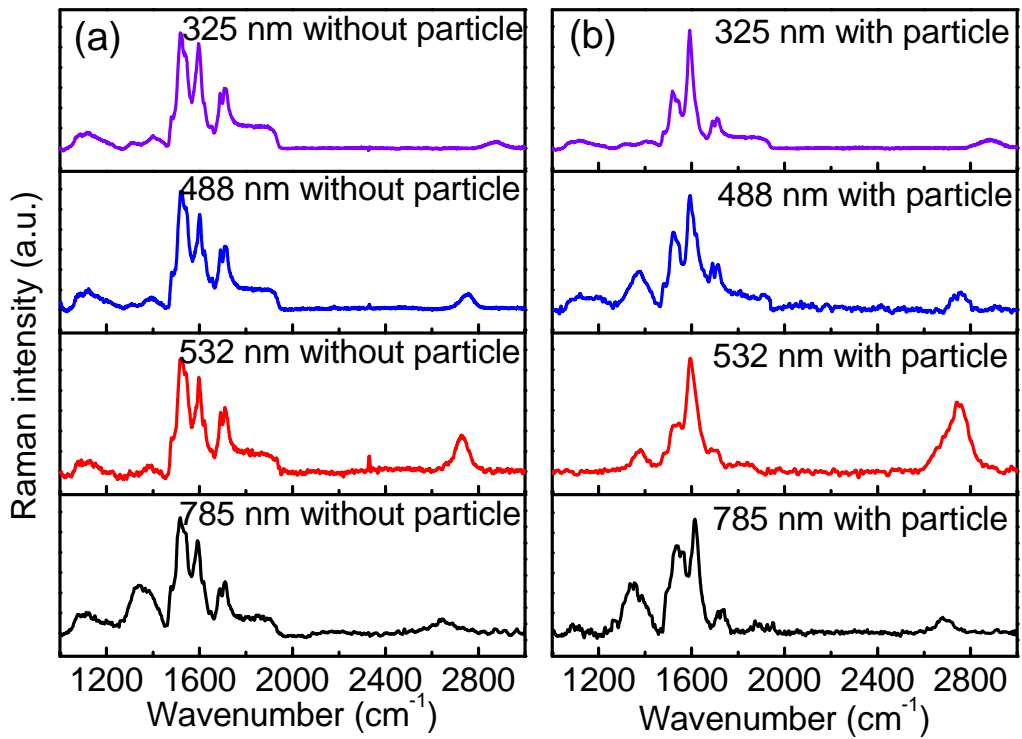
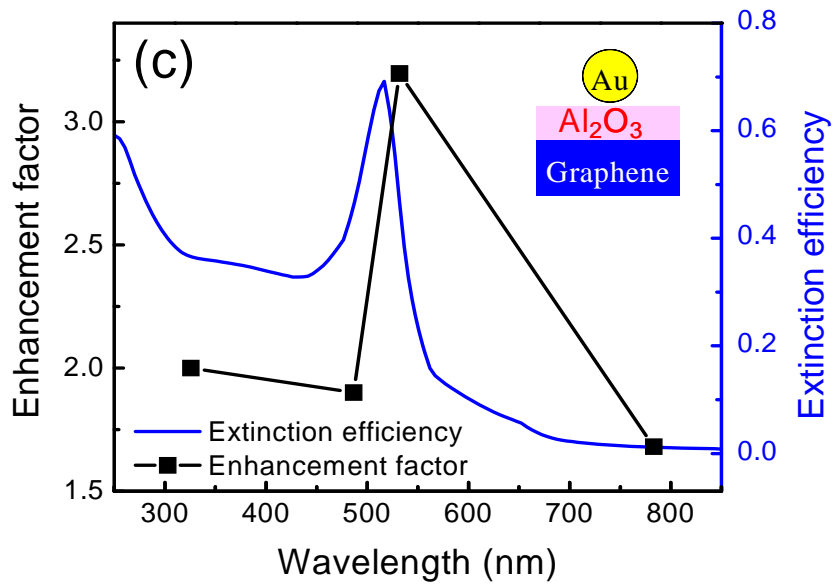

Figure 2

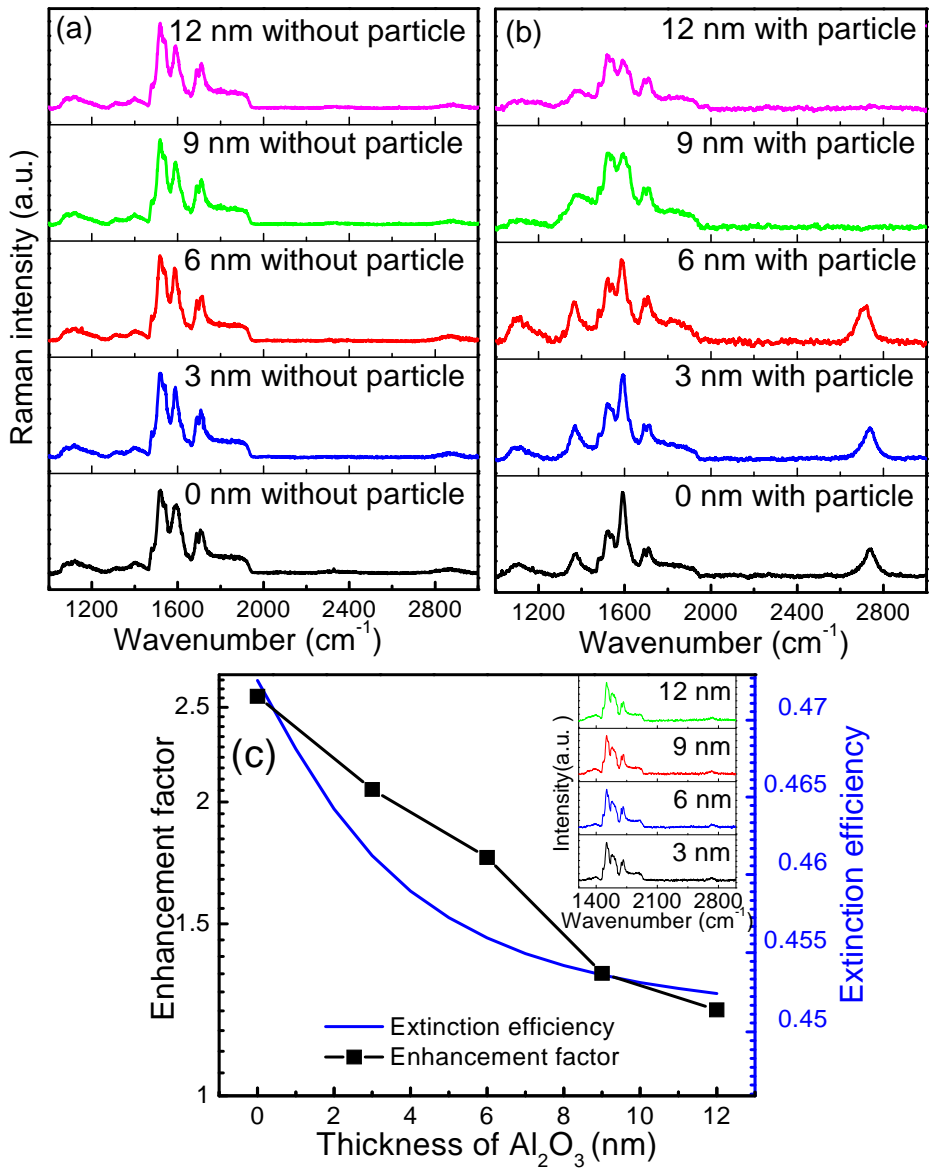

Figure 3